\pdfoutput=1
\documentclass[aps,pra,twocolumn,showpacs,groupedaddress]{revtex4}
\usepackage{graphicx}
\begin{document}

\title{Mitigation of loss within a molecular Stark decelerator}

\author{Brian C. Sawyer}
\email[]{sawyerbc@colorado.edu}
\author{Benjamin K. Stuhl}
\author{Benjamin L. Lev}
\author{Jun Ye}
\affiliation{JILA, National
Institute of Standards and Technology and University of Colorado \\
Department of Physics, University of Colorado, Boulder, CO
80309-0440, U.S.A.}
\author{Eric R. Hudson}
\affiliation{Department of Physics, Yale University, New Haven, CT
06520, USA}

\date{\today}

\begin{abstract}
The transverse motion inside a Stark decelerator plays a large role
in the total efficiency of deceleration. We differentiate between
two separate regimes of molecule loss during the slowing process.
The first mechanism involves distributed loss due to coupling of
transverse and longitudinal motion, while the second is a result of
the rapid decrease of the molecular velocity within the final few
stages. In this work, we describe these effects and present means
for overcoming them. Solutions based on modified switching time
sequences with the existing decelerator geometry lead to a large
gain of stable molecules in the intermediate velocity regime, but
fail to address the loss at very low final velocities. We propose a
new decelerator design, the quadrupole-guiding decelerator, which
eliminates distributed loss due to transverse/longitudinal couplings
throughout the slowing process and also exhibits gain over normal
deceleration to the lowest velocities.

\end{abstract}


\maketitle

\section{Introduction}
Recent development of cold polar-molecule sources promises to reveal
many interesting, and hitherto unexplored, molecular interaction
dynamics. The permanent electric dipole moment possessed by polar
molecules provides a new type of interaction in the ultracold
environment. This electric dipole-dipole interaction (and control
over it) should give rise to unique physics and chemistry including
novel cold-collision dynamics~\cite{Avdeenkov:2002,Hudson:PRA} and
quantum information processing~\cite{DeMille:QC}.

To date, cold polar-molecule samples have been produced most
successfully via three different mechanisms: buffer gas
cooling~\cite{Weinstein:CaHTrap,CampbellNH}; photo- and
magneto-association~\cite{PAReview:2006,RbCs:2005,KRb:2004,Ospelkaus:2006};
and Stark deceleration~\cite{Bethlem:1999}. Buffer gas cooling
achieves temperatures below 1 K through thermalization of molecules
with a He buffer. This technique produces relatively large densities
($10^{8}$ cm$^{-3}$) of polar ground-state molecules. However,
cooling below $\sim100$ mK has not yet been achieved because the He
buffer gas has not been removed quickly enough for evaporative
cooling~\cite{CampbellNH}. Photoassociation achieves the lowest
molecular temperatures of these techniques ($\sim$100 $\mu$K), but
is limited to molecules whose atomic constituents are amenable to
laser-cooling. Furthermore, molecules in their ground vibrational
state are not readily produced, yielding species with relatively
small effective electric dipoles, although this problem can be
overcome with more sophisticated laser control
techniques~\cite{RbCs:2005,Avi:2007}. Stark deceleration exists as
an alternative to these methods as the technique employs
well-characterized supersonic beam methods~\cite{Scoles} to produce
large densities of ground state polar molecules ($\sim$10$^{9}$
cm$^{-3}$), albeit at high packet velocities. One limitation of this
technique for trapping of decelerated molecules is an observed
drastic loss of slowed molecules at very low mean velocities in both
our own group's work and the Berlin group of G.
Meijer~\cite{BasPrivate}. We address this problem in this Article.

As the leading method for producing cold samples of chemically
interesting polar molecules, Stark deceleration has generated cold
samples of CO~\cite{Bethlem:1999}, ND$_3$~\cite{Bethlem:Trap},
OH~\cite{Bochinski:2003,Bochinski:2004,Hudson:2004,Bas:OH:2005},
YbF~\cite{Tarbutt:2004}, H$_2$CO~\cite{Hudson:PRA},
NH~\cite{Meerakker:NH2006}, and SO$_{2}$~\cite{Jung:2006}, leading
to the trapping of both ND$_3$~\cite{Bethlem:Trap} and
OH~\cite{Bas:OH:2005,Sawyer:2007}. Given the importance of Stark
deceleration to the study of cold molecules, it is crucial that the
technique be refined to achieve maximum deceleration efficiency. In
this work, we provide a detailed description of processes that limit
the efficiency of current decelerators and propose methods for
overcoming them. We propose possible solutions to the parametric
transverse/longitudinal coupling loss originally highlighted in
Ref.~\cite{meerakker:023401}, as well as elucidate a new loss
mechanism unique to producing the slowest molecules. We restrict the
described simulations and theory to Stark decelerated, ground-state
OH radicals, as the supporting experimental data was taken with this
molecular species. However, the loss mechanisms described herein are
not specific to OH, and represent a general limitation of current
Stark decelerators. This reduces the efficiency of Stark
decelerators for trapping cold polar molecules.

This Article is organized as follows. Section II describes the
mechanisms responsible for molecular loss at low velocities.
Sections III, IV, and V present methods of producing molecular
packets at intermediate velocities ($>$100 m/s) without the
distributed transverse/longitudinal coupling losses of
Ref.~\cite{meerakker:023401}. However, these methods exacerbate the
problem of low-velocity loss. Therefore, we propose a new
decelerator design in Section VI that exhibits gain over
conventional Stark deceleration to velocities as low as 14 m/s.

\section{Loss at Low Velocities}
In much of the previous work on Stark deceleration, it is assumed
that all motion parallel and transverse to the decelerator axis is
stable up to some maximum excursion velocity and position from the
beam center~\cite{Hudson:2004,Floris:2004}, enabling the derivation
of an analytical solution to predict stable-molecule phase-space
area. However, there are several important instances where the
assumptions of this model become invalid. In the case of very slow
molecules ($<$50 m/s) \footnote{This lower velocity limit depends on
the molecule of interest as well as decelerator electrode geometry.
In general, we expect this velocity limit to scale as
$\sqrt{\mu/m}$, where $\mu$ is the effective electric dipole moment
and $m$ is the mass of the given molecule.}, we observe two distinct
phenomena leading to reduced decelerator efficiency at the final
deceleration stages: transverse overfocusing and longitudinal
reflection. Transverse overfocusing occurs when the decelerated
molecules' speed becomes so low that the decelerating electrodes
focus the molecules too tightly (transversely) and they either make
contact with the electrodes or are strongly dispersed upon exiting
the decelerator. Longer decelerators tend to exacerbate this effect
due to the fact that molecules can travel at low speeds for many
stages. Nonetheless, there are several motivating factors for
constructing a longer decelerator. First, longer decelerators allow
less energy per stage to be removed and consequently lead to larger
longitudinal phase-space acceptance. Second, a longer decelerator
may allow deceleration of molecules possessing an unfavorable Stark
shift to mass ratio. We will discuss critical issues for use of such
long decelerators for slow molecule production.

A second low-velocity effect, which we have denoted ``longitudinal
reflection," is a direct result of the spatial inhomogeneity of the
electric field at the final deceleration stage. As highlighted in
the context of transverse guidance in Ref.~\cite{meerakker:023401},
the longitudinal potential is largest for those molecules
passing--in the transverse dimension--nearest to the decelerator
rods. However, the decelerator switching sequence is generally only
synchronous to a molecule on-axis traveling at the mean speed of the
packet. As a result, when the mean longitudinal kinetic energy of
the slowed packet becomes comparable to the Stark potential barrier
at the last stage, molecules off-axis can be stopped or reflected,
resulting in a spatial filtering effect. Furthermore, the
longitudinal velocity spread of the molecular packet at the final
stage, if larger than the final mean velocity, can lead to
reflection of the slowest portion of the packet. It is important
that the phenomena of overfocusing and longitudinal reflection be
addressed since molecule traps fed by Stark decelerators require
slow packets for efficient loading.

\begin{figure}
\begin{center}
\resizebox{0.8\columnwidth}{!}{\includegraphics{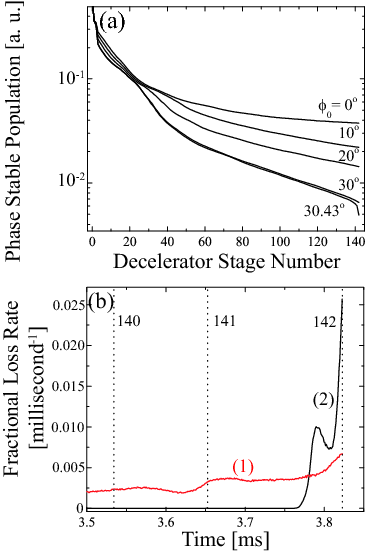}}
\caption{(color online) (a) Simulations of the phase-stable molecule
number as a function of stage number in our 142-stage decelerator.
Note the dramatic decrease in number in the last several stages for
$\phi_o = 30.43^\circ$. This decrease is due to transverse
overfocusing and longitudinal reflection of these slow (14 m/s)
molecules. (b) Simulated transverse (trace 1) and longitudinal
(trace 2) fractional loss rates as a function of time within the
final three stages at $\phi_o = 30.43^\circ$. The vertical dashed
lines denote the times of the given stage switches.} \label{CM2Loss}
\end{center}
\end{figure}

To illustrate these low-velocity effects, the number of phase-stable
molecules predicted by three-dimensional (3D) Monte Carlo simulation
is shown in Fig.~\ref{CM2Loss}(a) as a function of stage number for
increasing phase angle \footnote{All Monte Carlo simulation results
presented in this Article are based on three-dimensional models.}.
We define the deceleration phase angle, $\phi_{0}$, exactly as in
previous publications, where
$\phi_{0}=(z/L)180^{\circ}$~\cite{Hudson:2004}. The length of one
slowing stage is given by $L$ (5.5 mm for our machine), while the
molecule position between successive stages is denoted as $z$. We
define $z=0$ to be exactly between two slowing stages, therefore,
$\phi_{0}=0^{\circ}$ (bunching) yields no net deceleration. Phase
angles satisfying $0^{\circ}<\phi_{0}<90^{\circ}$ lead to
deceleration of the molecular packet, while the maximum energy is
removed for $\phi_{0}=90^{\circ}$. The 3D simulation results
displayed in Fig.~\ref{CM2Loss}(a) include both longitudinal and
transverse effects. The molecules have an initial velocity
($v_{initial}$) of 380 m/s, corresponding to the mean velocity of a
molecular pulse created via supersonic expansion in Xe. All
simulations and experimental data hereafter possess this
$v_{initial}$ unless otherwise noted.

As expected, a higher phase angle leads to a smaller number of
decelerated molecules. However, there is a sharp loss of molecules
in the last several deceleration stages for the highest phase of
$\phi_0 = 30.43^\circ$. This value of $\phi_{0}$ produces a packet
possessing a final velocity ($v_{final}$) of 14 m/s. This loss is
attributed to transverse overfocusing and longitudinal reflection.
These distinct effects are illustrated in Fig.~\ref{CM2Loss}(b),
which displays the transverse (trace 1) and longitudinal (trace 2)
fractional loss rate of molecules traversing the final three slowing
stages at $\phi_0 = 30.43^\circ$, $v_{final}=14$ m/s. The switching
time for each stage is denoted by a vertical dashed line, which is
labeled by the corresponding stage number. Longitudinal reflection
of molecules is clearly the dominant loss mechanism for the lowest
final velocity shown in Fig.~\ref{CM2Loss}(a). Nonetheless, there
also exists a non-negligible rise in transverse losses at the final
stage. That is, because the molecular beam is moving very slowly in
the last few deceleration stages, the transverse guiding fields of
the decelerator electrodes have a greater focusing effect on the
molecules (see Eq. 3 of Ref.~\cite{Bochinski:2004}) and focus the
molecules so tightly that they collide with a deceleration stage and
are lost. In the case of our decelerator, this leads to loss of 20\%
of the decelerated molecule number between $\phi_o = 30^\circ$ (50
m/s) and $\phi_o = 30.43^\circ$ (14 m/s). Such a dramatic loss is
not predicted by analytical theory, as the stable phase-space area
decreases by only $<$1\% over this range of $\phi_0$. This number is
calculated directly after the decelerator is switched-off and is
thus an upper bound, since experiments employing these cold
molecules require them to travel out of the decelerator where
transverse spread can lead to dramatic loss of molecule number.

\begin{figure}
\begin{center}
\resizebox{1\columnwidth}{!}{
    \includegraphics{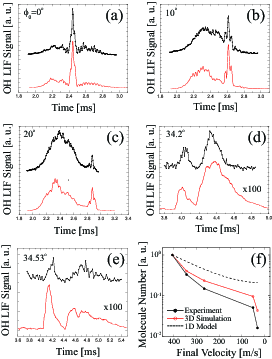}%
}
\end{center}
\caption{(color online) (a-e) Experimental ToF profiles (upper
curves) with Monte Carlo simulation results (lower curves) for
slowing at various phase angles for $v_{initial}=406$ m/s. The
vertical axes of panels (d) and (e) have been scaled up by a factor
of 100. Note the good correspondence of simulation and data to
velocities as low as 30 m/s ($\phi_o = 34.53^{\circ}$). (f)
Experimental (dots) and simulated (open circles) total molecule
number as a function of final velocity for the data in panels (a-e).
The dashed curve is the expected decelerator efficiency calculated
from the one-dimensional (1D) theoretical model of
Ref.~\cite{Hudson:2004}. \label{ToFS=1}}
\end{figure}

Experimental evidence of this sudden decrease in molecule number at
very low velocities is given in Fig.~\ref{ToFS=1}, which displays
time-of-flight (ToF) spectra along with corresponding Monte Carlo
simulation results at various phase angles. The decelerated OH
molecules are in the weak-field seeking $|F=2,m_{F}=\pm2,-\rangle$
state. The first two quantum numbers of the state denote its
hyperfine components, while the third number indicates the parity of
the state in the absence of electric fields. Figures~\ref{ToFS=1}(b)
and (c) illustrate progressive time delays of the slowed molecular
packets from the background of non-synchronous molecules.
Figures~\ref{ToFS=1}(d) and (e) display only the decelerated
packets, and the vertical axes of these plots are magnified by a
factor of 100. The dual peaks correspond to molecular packets loaded
into two successive stages, and the later peaks represent the
intended final velocities. The total detected and simulated molecule
numbers are plotted in Fig.~\ref{ToFS=1}(f) as a function of final
velocity, along with the theoretically expected decelerator
efficiency (dashed line)~\cite{Hudson:2004}. Note that the sudden
population decrease in both simulation and experimental results is
not reflected in the one-dimensional theoretical model, which does
not account for the transverse dynamics or field inhomogeneities
that cause such behavior. This effect is detrimental to the
production of dense samples of cold molecules.

To remove the overfocusing effect at low velocities, the transverse
focusing of the last several decelerator stages needs to be reduced.
Different types of transverse focusing elements may be inserted into
the deceleration beam line to compensate for this phenomenon. This
idea is discussed in detail in Sections IV-VI. We note that the
proposed solutions, while successful in addressing the detrimental
longitudinal/transverse coupling effects, do not mitigate the
problem of longitudinal reflection at low velocities.

\section{Distributed Loss}
\begin{figure}
\begin{center}
\resizebox{1\columnwidth}{!}{
    \includegraphics{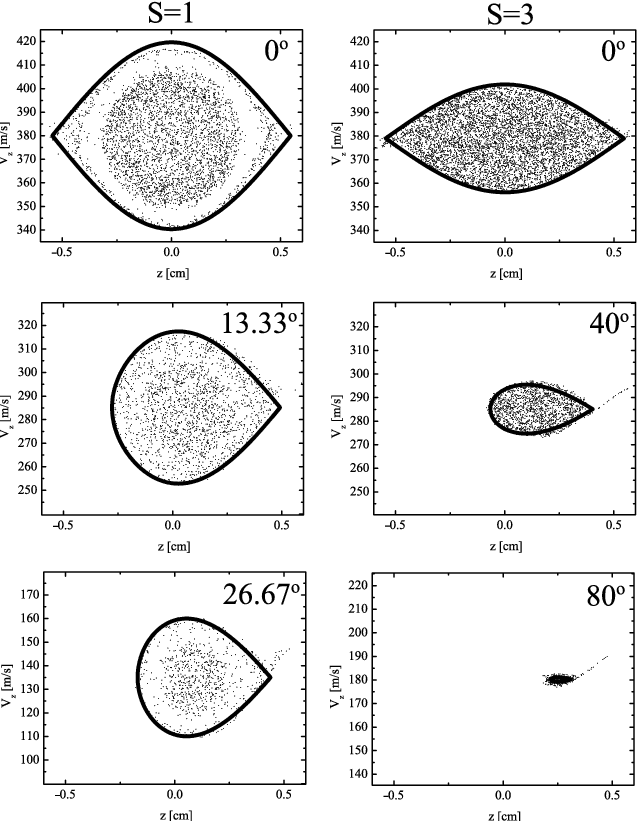}%
}
\end{center}
\caption{Monte Carlo simulation results for the longitudinal phase
space of decelerated molecules. The left column shows $\phi_o$ =
0$^\circ$, $13.33^\circ$, and $26.67^\circ$ for S = 1 slowing, while
the right column shows $\phi_o$ = 0$^\circ$, $40^\circ$, and
$80^\circ$ for S = 3 deceleration. The factor of three between $S=1$
and $S=3$ phase angles ensures that molecules have roughly the same
final velocities. The observed velocity difference at high phase
angles occurs because the 142 stages of our slower is not a multiple
of three. Note that, although the $S=3$ phase plot is more densely
populated than that of $S=1$ at $\phi_o$ = 0$^\circ$, its
phase-space acceptance decreases dramatically relative to $S=1$ at
the lowest velocities. All plots are generated using an identical
initial number of molecules, and therefore the density of points is
meaningful for comparison. For definition of $S=1$ and $S=3$, see
Ref.~\cite{meerakker:023401} and Section IV. \label{PSCompare}}
\end{figure}

Coupling between transverse and longitudinal motion throughout the
deceleration sequence invalidates the assumptions made in
Ref.~\cite{Hudson:2004,Floris:2004}. The fact that the transverse
guidance of the molecular beam comes from the same electrodes that
provide the deceleration means that the longitudinal and transverse
motions are necessarily coupled. While this phenomenon is well
understood in the field of accelerator physics~\cite{Wiedemann}, it
was first pointed out in the context of Stark deceleration in
Ref.~\cite{meerakker:023401}. The result can be seen in the left
column of Fig.~\ref{PSCompare}, where the longitudinal phase space
of OH packets is shown versus increasing phase angle. In
Fig.~\ref{PSCompare}, the dark lines represent the separatrix,
partitioning stable deceleration phase space from that of unstable
motion as calculated from Eq. 2 in Ref.~\cite{Hudson:2004}. Each dot
represents the position in phase space of a simulated molecule. In
the absence of coupling between the longitudinal and transverse
motions, one would expect the entire area inside the separatrix to
be occupied. Therefore, the structure in these graphs is evidence of
the importance of the transverse motion.

In the left column of Fig.~\ref{PSCompare}, the coupling of
longitudinal and transverse motions is responsible for two effects
\footnote{For these phase space simulations, the input molecular
beam has longitudinal spatial and velocity distributions that
overfill the acceptance area.}: First, in the center of the stable
area at $\phi_{0}=0^{\circ}$--near the synchronous molecule--the
density of stable molecules is less than in the surrounding area.
This is because molecules that oscillate very near the synchronous
molecule experience little transverse guiding. This effect is not
dramatic and is only discernable for an exceedingly large number of
stages. Furthermore, this effect is even less important for the
increased phase angles typically used for deceleration, since for
these switching sequences the synchronous molecule experiences more
of the transverse guiding forces than it does during bunching. The
second effect, which is much more evident, is the absence of
molecules at intermediate distances from the synchronous molecule as
shown in the left column of Fig~\ref{PSCompare}. This so-called
`halo' is due to parametric amplification of the transverse motion
and is similar to the effects seen in cold molecule storage
rings~\cite{Floris:2004} as well as charged particle
accelerators~\cite{Suzuki:1985}. Essentially, the longitudinal
oscillation frequency of a molecule in this region is matched to the
transverse oscillation frequency, leading to amplification of the
transverse and longitudinal motion and consequent loss.

\begin{figure}
\begin{center}
\resizebox{0.8\columnwidth}{!}{
    \includegraphics{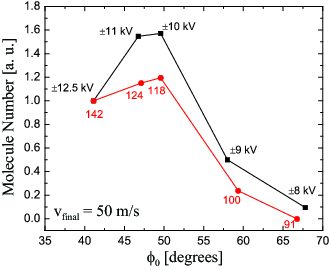}%
}
\end{center}
\caption{(color online) Experimental results from changing the
voltage on decelerator rods (squares) and decreasing the effective
decelerator length (circles). Effective slower length is modified by
initially operating the decelerator at $\phi_{0}=0^{\circ}$, $S=3$,
then slowing with $S=1$ to $v_{final}=50$ m/s for the number of
stages labeled. Both curves illustrate that transverse/longitudinal
couplings are strongly dependent on phase angle, and have a marked
effect on decelerator efficiency. \label{VLChange}}
\end{figure}

There is a compromise between decreasing longitudinal phase-space
area and increasing transverse guidance for increasing $\phi_{0}$.
To demonstrate this effect, we decelerate molecules to a fixed
$v_{final}$ and vary the phase angle used to reach this velocity.
This is done either by changing the voltage applied to the
decelerator rods or by modifying the effective length of the
decelerator itself for each $\phi_{0}$ of interest. The experimental
data shown in Fig.~\ref{VLChange} is the result of varying the
voltage applied to our decelerator rods (squares) and the effective
decelerator length (circles). Both lowering the decelerator voltage
and using shorter lengths of the decelerator for slowing requires
increasing $\phi_{0}$ to observe the same $v_{final}$ of 50 m/s. We
are able to effectively shorten the decelerator by initially
bunching the packet for a given number of stages before beginning a
slowing sequence. Note that we use $S=3$ bunching to remove any
transverse/longitudinal couplings during these first stages, then
switch back to $S=1$ slowing for the remainder of the decelerator.
The parameter $S$ signifies the mode of decelerator operation and is
previously defined in Ref.~\cite{meerakker:023401} as well as in
Section IV. The phase stable region of $S=1$ for
$\phi_{0}\geq40^{\circ}$ is entirely contained within that of $S=3$
at $\phi_{0}=0^{\circ}$, therefore no artifacts from initial
velocity filtering are present in this data. We observe that,
contrary to the predictions of the one-dimensional
theory~\cite{Hudson:2004}, a higher phase angle can lead to greater
decelerator efficiency up to some maximum $\phi_{0}$. This is a
direct consequence of distributed transverse/longitudinal couplings
illustrated in Fig.~\ref{PSCompare}. At even larger phase angles,
the longitudinal phase-space acceptance becomes a limiting factor.
The labels next to each data point correspond to either the voltage
applied to the decelerator rods (squares) or the number of utilized
$S=1$ slowing stages (circles). Figure~\ref{VLChange} further
illustrates that the transverse/longitudinal couplings outlined by
Ref.~\cite{meerakker:023401} reduce decelerator efficiency, and are
highly dependent on phase angle.

The coupling between longitudinal and transverse motion is
detrimental to efficient operation of a Stark decelerator, reducing
the total number of decelerated molecules. This effect will be even
worse for decelerating molecules with an unfavorable Stark
shift-to-mass ratio. Fortunately, the transverse and longitudinal
motions can be decoupled by introducing a transverse focusing
element to the deceleration beam line that overwhelms the transverse
focusing provided by the deceleration electrodes in analogy to the
focusing magnets of charged-beam machines. This technique also has
the advantage of providing a larger stable region in the transverse
phase space which further enhances the decelerated molecule number.
The remainder of this manuscript discusses methods of implementing a
transverse focusing element to decouple the longitudinal and
transverse motion. Sections IV and V describe methods that can be
implemented by modifying the timing sequences of present
decelerators with moderate success at intermediate $v_{final}$.
Section VI presents an improved design for a Stark decelerator that
solves this problem of distributed longitudinal and transverse loss
and also reduces the previously described overfocusing losses at the
final stage.

\section{Decelerator Overtones}
\begin{figure}
\begin{center}
\resizebox{0.9\columnwidth}{!}{
    \includegraphics{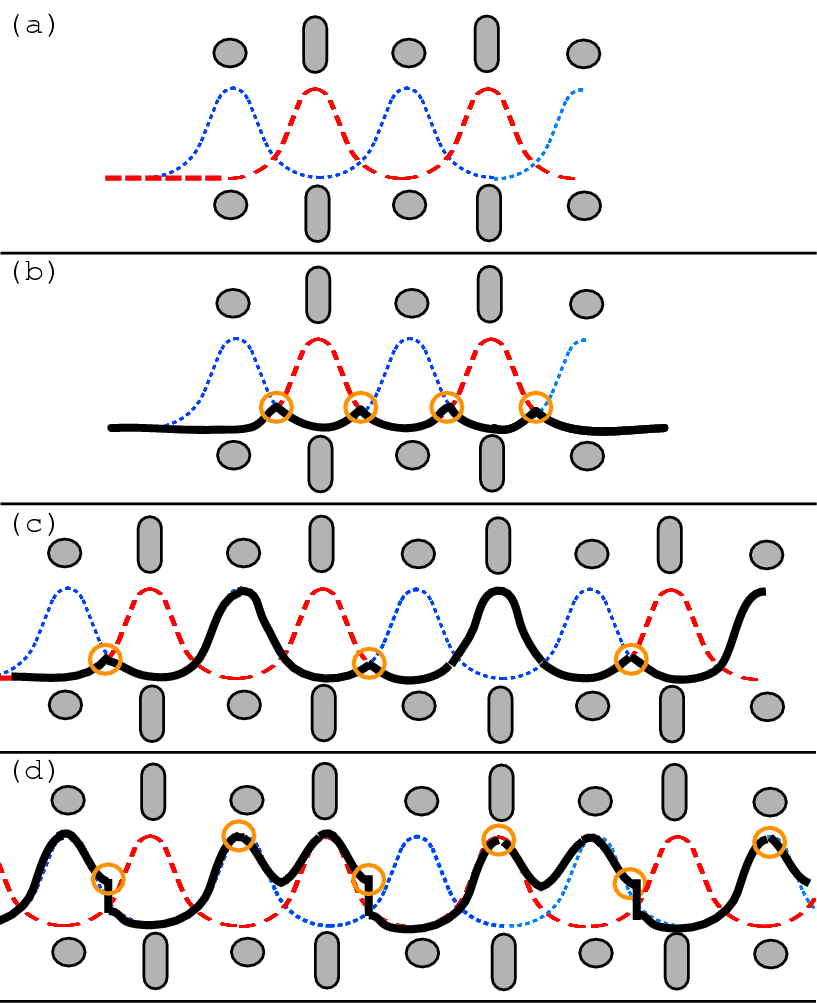}%
}
\end{center}
\caption{(color online) Deceleration schemes. (a) Potential energy
shift of polar molecules in the Stark decelerator. The dotted (blue)
curves show the potential energy shift when the horizontal (circular
cross section) electrodes are energized, while the dashed (red)
curves show the potential energy shift when the vertical (elongated
cross section) electrodes are energized. Deceleration proceeds by
switching between the two sets of energized electrodes. In panels
(a)-(c), the thick black line indicates the potential experienced by
the molecules. The empty circles indicate a switching event. (b)
Traditional (S = 1) operation at $\phi_o$ = 0$^\circ$. For phase
stability, the switching always occurs when the molecules are on an
upward slope, and as such the molecules are never between a pair of
energized electrodes. Thus, the maximum transverse guiding is never
realized. (c) First overtone operation ($S = 3$) at $\phi_o$ =
0$^\circ$. By switching at one-third of the $S = 1$ rate, the
molecules are allowed to fly directly between an energized electrode
pair, and thus, experience enhanced transverse guiding. (d)
Optimized first overtone operation ($S = 3+$) at $\phi_o$ =
0$^\circ$: Initially, the packet rises the Stark potential created
by one set of electrodes. When the molecules reach the apex of this
potential, the alternate set of electrodes is energized in addition.
In this way, the molecules experience one more stage of maximum
transverse guiding for each slowing stage. Note that, to minimize
the un-bunching effect, the grounded-set of electrodes is switched
on when the molecules are directly between the energized electrodes.
\label{DecelerationSchemes}}
\end{figure}

The simplest method for introducing a transverse focusing element to
the decelerator beam line is to let the molecules fly through an
energized deceleration stage without removing the field. In this
manner, molecules experience the transverse focusing of the entire
stage without their longitudinal motion affected. Traditional
longitudinal phase stability requires the switching of the fields to
occur on an upward slope of the molecular potential energy, i.e.,
faster molecules are slowed more while slower molecules are slowed
less than the synchronous molecule. Hence, it is necessary to
de-sample the bunching switching rate by an odd factor (3,5,7...):
the so-called decelerator overtones~\cite{meerakker:053409}. For
convenience, we define the quantity $S= v_o/v_{Switch}$, where $v_o$
is the synchronous molecule velocity and the switching speed
$v_{Switch}$ is given as the stage spacing $L$ divided by the
switching time-interval. Reference~\cite{meerakker:053409}
considered only the bunching case. In this work, we generalize to
the case of actual deceleration. However, the above definition of
$S$ is still valid. That is, $S$ is constant despite the fact that
both $v_o$ and $v_{Switch}$ vary when $\phi_o
> 0^\circ$. With this definition we see that traditional
deceleration can be described by $S = 1$, while the method of
de-sampling the switch rate by a factor 3 is described by $S = 3$.
These two methods of deceleration can be seen in
Figs.~\ref{DecelerationSchemes} (b) and (c), where their respective
switching schemes are shown for $\phi_o = 0^\circ$. By switching at
one-third the rate, the molecule packet flies through a deceleration
stage that is energized and experiences enhanced transverse guiding.

Longitudinal phase space simulations of $S = 3$ slowing at various
phase angles are shown in the right column of Fig.~\ref{PSCompare}.
No structure is present in these plots. Also, the region of
longitudinal phase stability for $S = 3$--even at
$\phi_{0}=0^{\circ}$--is reduced compared to $S = 1$. This is
because the maximum stable velocity, as calculated from Eqs. 2 and 6
of Ref.~\cite{Hudson:2004}, depends on the spacing between
deceleration stages as $L^{-1/2}$, and thus, the separatrix velocity
bound is reduced by a factor of $\sqrt{3}$ \footnote{Physically,
this is because the molecules fly longer between deceleration
stages, and thus, the velocity mismatch can lead to a larger
accumulation of spatial mismatch.}. Nonetheless, the absence of
coupling to the transverse motion leads to a larger number of
molecules for the $\phi_o = 0^\circ$ case shown in the uppermost
panel of Fig.~\ref{PSCompare}. In a given decelerator, $S=3 $
slowing requires a factor of three higher phase angle than $S=1$ to
reach the same final velocity. As illustrated in
Fig.~\ref{PSCompare}, this fact severely limits the practicality of
$S=3$ as a deceleration scheme, as it implies a dramatic reduction
in velocity acceptance at the highest phase angles.

\begin{figure}
\begin{center}
\resizebox{1\columnwidth}{!}{
    \includegraphics{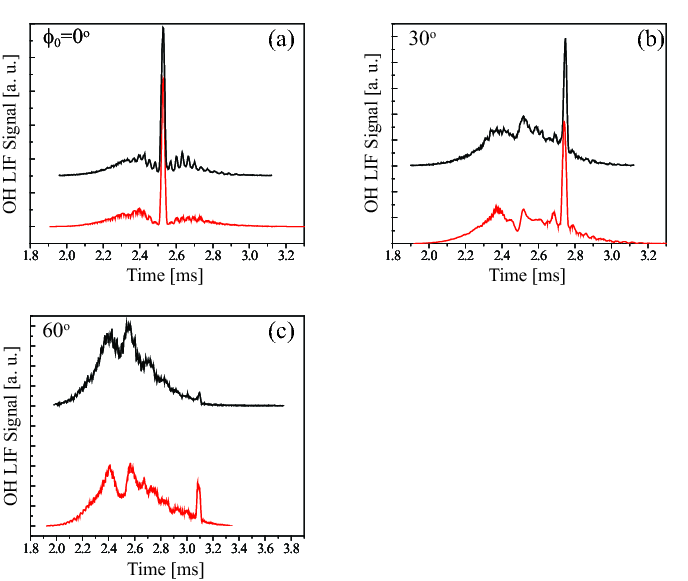}%
}
\end{center}
\caption{\label{ToFS3} (color online) Experimental ToF data (upper
curves) of OH molecules decelerated using S = 3 at varying phase
angles. The results of 3D Monte Carlo simulations are also shown
(lower curves).}
\end{figure}

Figure~\ref{ToFS3} shows ToF data (upper curves) taken in our
142-stage decelerator for $S = 3$ with increasing phase angle along
with the results of 3D Monte Carlo simulations (lower curves).
Qualitatively, the ToF signal is no different than for $S = 1$.

\begin{figure}
\begin{center}
\resizebox{1\columnwidth}{!}{
    \includegraphics{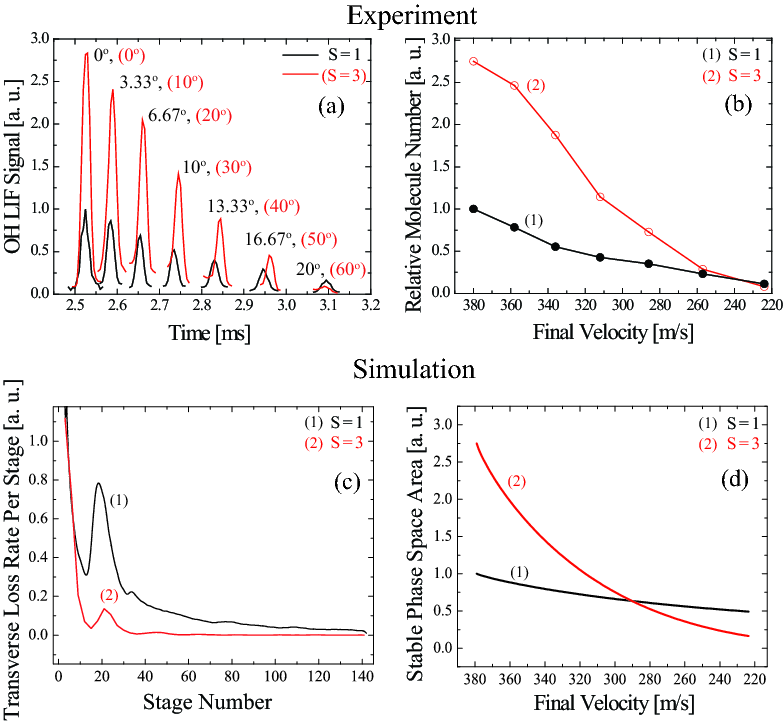}%
}
\end{center}
\caption{\label{S3S1Compare}(color online) Comparison of
deceleration using $S = 3$ versus $S=1$. (a) Experimental ToF data
of decelerated OH packets with $S=3$ (top) and $S=1$ (bottom). Note
the factor of three between $S=1$ and $S=3$ phase angles. (b)
De-convolved, integrated molecule number for $S=3$ (trace 2) and
$S=1$ (trace 1) for the packets shown in panel (a). (c) Simulated
transverse loss rate per stage for $S=1$ (trace 1) and $S=3$ (trace
2) deceleration. As expected, $S=1$ results in larger transverse
loss rates throughout. (d) Calculated stable longitudinal
phase-space area for $S=1$ (trace 1) and $S=3$ (trace 2), with
initial points scaled to the experimental ratio of $2.75$. The above
panels highlight that the observed shortcoming of $S=3$ deceleration
is entirely due to loss of longitudinal velocity acceptance at high
phase angles.}
\end{figure}

Shown in Fig.~\ref{S3S1Compare}(a) are experimental decelerated
molecular packets for $S = 3$ (upper) and $S = 1$ (lower) versus
increasing phase angle. For each successive packet, the $S = 3$
phase angle increases by $10^\circ$ in the range $\phi_o =
0^\circ$--60$^{\circ}$, while the $S = 1$ phase angle increases by
$10^\circ/3$. In this manner, the packets are decelerated to roughly
the same velocity. There is a slight difference at the highest phase
angles shown because the total number of stages in our decelerator
(142) is not an exact multiple of 3. In Fig.~\ref{S3S1Compare}(b)
the de-convolved total molecule number for each of these packets is
plotted versus final speed. While the $S = 3$ method dominates over
$S = 1$ for small phase angles, its effectiveness decreases as the
deceleration becomes more aggressive--by 224 m/s, the $S = 1$
molecule number is already larger than that of $S = 3$. This
behavior is expected since the phase angle used to decelerate to 224
m/s is $60^\circ$, while the required $S = 1$ phase angle is only
$20^\circ$. Although the $S = 3$ longitudinal phase bucket does not
exhibit structure, it is so much smaller in enclosed area than the
$S = 1$ that its total molecule number is smaller. The simulation
results of Fig.~\ref{S3S1Compare}(c) and the theory results of
Fig.~\ref{S3S1Compare}(d) support this description.
Figure~\ref{S3S1Compare}(c) shows that, even at $\phi_o = 80^\circ$,
the transverse loss rate per stage is at all times greater for $S=1$
than $S=3$. However, the calculated longitudinal phase-space
acceptance of Fig.~\ref{S3S1Compare}(d) mirrors the behavior
observed experimentally in Fig.~\ref{S3S1Compare}(b) when the
initial points are scaled to the experimental ratio of 2.75. This
scaling accounts for the aforementioned `halo' in the slowed $S=1$
packet, which persists relatively unchanged over the range of $S=1$
phase angles used ($\phi_o = 0^\circ-20^\circ$). The fact that the
theory curves of Fig.~\ref{S3S1Compare}(d) cross at a higher
velocity than the data of Fig.~\ref{S3S1Compare}(b) suggests there
is increased transverse guiding of $S=3$ slowing at high phase
angles. Nevertheless, even with 142 stages of deceleration, $S=3$ is
unfavorable for velocities below 224 m/s due to reduced stable
phase-space area.

\begin{figure}
\begin{center}
\resizebox{0.8\columnwidth}{!}{
    \includegraphics{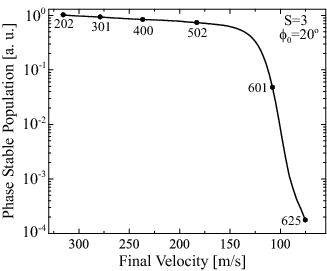}%
}
\end{center}
\caption{\label{S=3StageNumber}Monte Carlo results for decelerated
molecule number using $S = 3$ and $\phi_o$ = 20$^\circ$ versus final
velocity. The number next to each data point is the number of stages
used. Because of transverse overfocusing and longitudinal velocity
filtering, essentially no molecules survive below 100 m/s. }
\end{figure}

At this point, one would expect that operation at lower phase angles
would permit realization of the gain produced by the $S=3$ method.
This can be accomplished by naively using a longer decelerator. A
simulation of this kind is presented in Fig.~\ref{S=3StageNumber},
which plots the number of molecules present after deceleration at
$\phi_o = 20^\circ$ versus final velocity \footnote{In this
simulation, $\phi_o = 20^\circ$ is chosen because it produces the
most gain over $S = 1$ in our deceleration experiments.}. The number
next to a data point represents the number of deceleration stages
used. Initially the decelerated molecule number is relatively flat
versus final velocity. However, after about 500 stages (180 m/s),
the number of decelerated molecules begins to decrease and
dramatically falls off after 550 stages (150 m/s). Very few
molecules survive below 100 m/s. This is because for $S = 3$ the
decelerated molecules must fly through an entire stage while
experiencing a guiding force in only one dimension [see
Fig.~\ref{DecelerationSchemes}(c)]. Once the molecules are at slower
speeds they can spread out in one transverse dimension or be
over-focused in the other and collide with the rods. As the mean
kinetic energy of the slowed packet becomes comparable to the full
potential height, the packet can be nearly stopped as it traverses
the intermediate charged stage. This has two consequences: (1)
longer transit time leading to more intense transverse focusing; and
(2) velocity filtering of the low-speed packet as slower molecules
are longitudinally reflected from this potential.

The transverse loss responsible for the extreme drop in molecule
number for $S = 3$ deceleration also occurs in traditional
deceleration, but to a lesser degree. Because of this decrease in
molecule number at low speeds, the usefulness of slowing with $S =
3$ is generally limited to experiments that do not require the
lowest velocities, such as microwave spectroscopy and collision
experiments~\cite{Hudson:alphadot,Lev:OHspec,Joop:2006}. We note
also that our simulations predict no low-velocity gain when using
slowing sequences containing combinations of deceleration at $S=1$
and bunching at $S = 3$.

\section{Modified Decelerator Overtones}
\begin{figure}
\begin{center}
\resizebox{0.8\columnwidth}{!}{
    \includegraphics{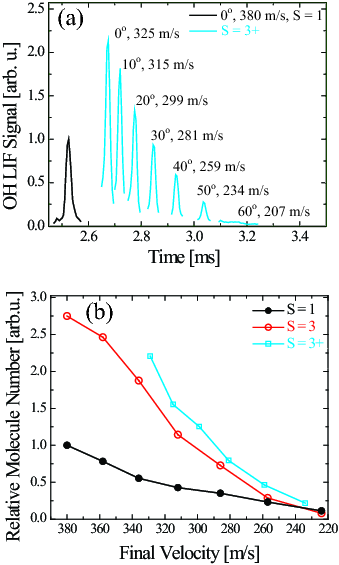}%
}
\end{center}
\caption{(color online) (a) Experimental ToF data of decelerated OH
packets produced using the $S = 3+$ modified overtone. Also shown
for comparison is the experimental bunching packet for operation at
$S = 1$. (b) The de-convolved, integrated molecule number calculated
from $S = 3+$ (open squares), $S = 3$ (open circles), and $S = 1$
(dots) data.\label{S=3+Loading}}
\end{figure}

A natural extension of the above overtone deceleration is the use of
what we have termed a ``modified decelerator overtone," denoted by
an additional plus sign, i.e., $S = 3+$. Deceleration in this manner
is shown in Fig.~\ref{DecelerationSchemes}(d) for $\phi_o =
0^\circ$. In this method, deceleration proceeds similarly to
conventional $S = 3$. However, $S=3+$ sequences yield confinement of
the packet in both transverse dimensions. This is achieved by
charging all slower rods for the period in which the synchronous
molecule is between switching stages. In order to minimally disrupt
the longitudinal dynamics of the synchronous molecule, the second
set of slower rods is charged only when the molecule is at the peak
of the longitudinal potential from the first rod set. The packet
then traverses two charged rod pairs before reaching the next
switching stage, at which point the rod pair that was originally
charged is grounded. While this does create a slight anti-bunching
effect, i.e., molecules in front of the synchronous molecule gain a
small amount of energy, it provides an extra stage of transverse
guidance in comparison to $S = 3$. Experimental results of this
method of slowing are shown in Fig.~\ref{S=3+Loading} for comparison
to deceleration using both $S = 1$ and $S = 3$.
Fig.~\ref{S=3+Loading}(a) displays a unique consequence of the
$S=3+$ switching sequence,  where operation at $\phi_o$ = 0$^\circ$
leads to deceleration. Figure~\ref{S=3+Loading}(b) shows that
operation with $S = 3+$ provides slightly more molecules than $S =
3$, but the loss of molecules due to decreased longitudinal velocity
acceptance remains a problem. The increase in molecule number for $S
= 3+$ over $S = 3$ is due to the extra stage of transverse guidance,
which for the higher-velocity packets we measured leads to a larger
transverse acceptance.

\begin{figure}
\begin{center}
\resizebox{0.8\columnwidth}{!}{
    \includegraphics{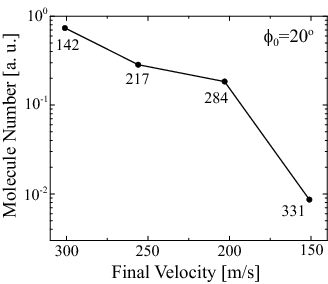}%
}
\end{center}
\caption{Monte Carlo results of decelerated molecule number using $S
= 3+$ and $\phi_o$ = 20$^\circ$ versus final velocity. The number
next to each data point is the number of stages used. Because of
transverse overfocusing, essentially no molecules survive below 100
m/s. \label{S=3+Number}}
\end{figure}

To determine whether the extra stage of transverse guidance would
counter the overfocusing effects \footnote{While this may seem
counterintuitive, in some cases when the transverse overfocusing is
not too strong, the addition of another focusing element can change
the sign of the molecules transverse velocity, keeping the beam
confined within the decelerator.}, we perform simulations of $S =
3+$ deceleration at $\phi_o$ = 20$^\circ$ for a varying number of
deceleration stages. The results of these simulations, shown in
Fig.~\ref{S=3+Number}, are similar to the results for $S = 3$.
Namely, as the decelerator length is increased and the molecules'
speed is reduced, there is a marked molecule number loss for
velocities below 200 m/s. In our simulations, we could not observe
any molecules below 100 m/s. Again, transverse overfocusing and
longitudinal filtering of the molecular packet by the deceleration
electrodes are responsible for large losses in the decelerator, and
this suggests that measures beyond modified switching schemes are
required to overcome these loss mechanisms.

\begin{figure}
\begin{center}
\resizebox{0.8\columnwidth}{!}{
    \includegraphics{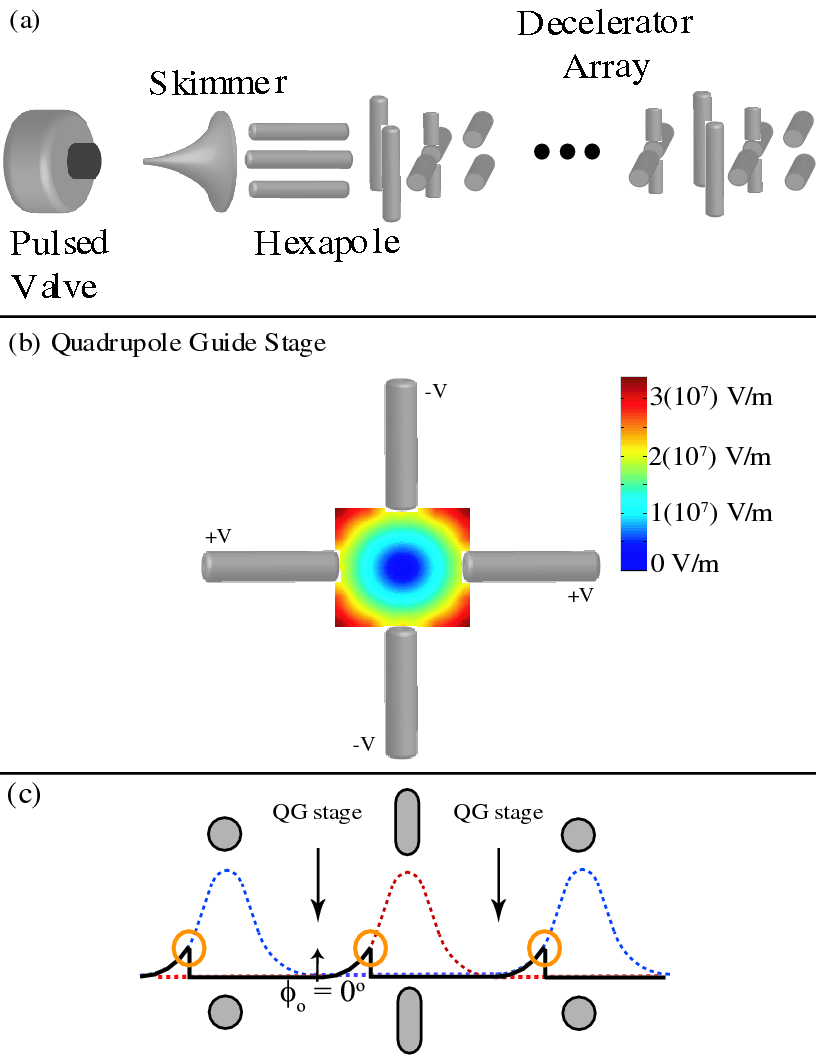}
}
\end{center}
\caption{(color online) Quadrupole-guiding decelerator. (a)
Schematic of QGD. (b) Electric field of quadrupole guiding stage
energized to $\pm$12.5 kV. (c) Switching scheme for deceleration
with the QGD. \label{AlternativeSDSchematic}}
\end{figure}

\section{Quadrupole-Guiding Decelerator}

In addition to modifying the timing sequences of existing
decelerators, it is possible, and perhaps preferable, to uncouple
the longitudinal motion inside the Stark decelerator from the
transverse motion by redesigning the decelerator electrode geometry.
One simple redesign, which we call the quadrupole-guiding
decelerator (QGD), is shown in Fig.~\ref{AlternativeSDSchematic}(a).
In this decelerator, a quadrupole-guiding stage
(Fig.~\ref{AlternativeSDSchematic}(b)) is interleaved between each
deceleration stage. While it may not be necessary to have a
quadrupole-guiding stage between each deceleration stage (especially
in the beginning of the decelerator), it simplifies the analysis and
will be used here. The switching of the electric fields inside a QGD
is similar to a traditional decelerator operated with $S =1$.
Figure~\ref{AlternativeSDSchematic}(c) shows the potential energy
experienced by a molecule decelerated at $\phi_o \approx 45^\circ$
and is represented by a thick black curve. In this panel, the
quadrupole-guiding electrodes are omitted for clarity. Note that the
quadrupoles are always energized and their center coincides with the
$\phi_o = 0^\circ$ position, while $\phi_o = 90^\circ$ occurs
between the deceleration electrodes.

\begin{figure}
\begin{center}
\resizebox{0.8\columnwidth}{!}{
    \includegraphics{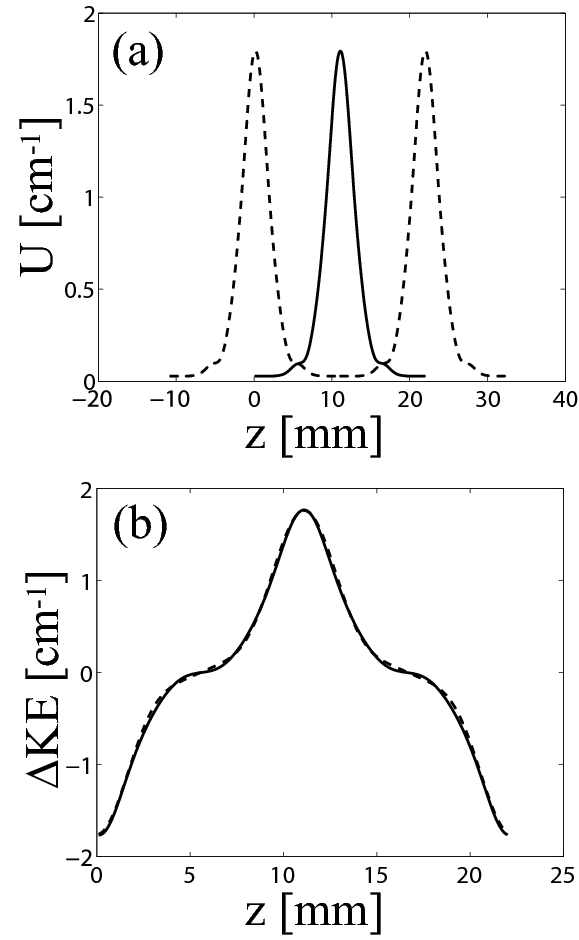}%
}
\end{center}
\caption{(a) The Stark shift of an OH molecule in the
$|2\pm2,-\rangle$ state inside the QGD. The solid curve is the Stark
shift due to the slowing electrodes, while the dashed curve is the
Stark shift due to the electrodes that will be energized at the
switching time. (b) The change in the molecule's kinetic energy as a
function of position is shown (solid) as well as a fit of
Eq.~\ref{DeltaKEFit}, including up to n = 3 (dashed). The solid
curve is calculated from the subtraction of the two curves in panel
(a). \label{QGDLongForces}}
\end{figure}

Deceleration with a QGD enjoys the same longitudinal phase-stability
as a traditional decelerator. In a QGD, the maximum stable excursion
position $\Delta \phi_{max}$ and velocity $\Delta v_{max}$ will
differ from that of the traditional decelerator because the
decelerating electrodes are most likely further apart. In other
words, to prevent high-voltage breakdown, the quadrupole stages
require the same inter-stage spacing as deceleration stages in a
traditional decelerator. Thus, the decelerating electrodes for a QGD
will be twice as far apart as in our traditional decelerator, which
possesses an inter-stage spacing of $5.5$ mm. Since the dependence
of the decelerating force on $\phi_o$ is less steep, the shape of
the stable longitudinal phase space will change. Understanding the
shape of the stable longitudinal phase area is crucial for
predicting the performance of the QGD, and can be derived by
examining the longitudinal forces inside the QGD. Shown in
Fig.~\ref{QGDLongForces}(a) is the on-axis Stark shift of an OH
molecule in the $|2,\pm2,-\rangle$ state inside the unit cell,
defined as 3 deceleration stages of the QGD. The solid line is the
Stark shift due to the slowing stage centered at 11 mm, while the
dashed line is the Stark shift of the stages which will be energized
when the fields are switched. The subtraction of these two curves,
shown in Fig.~\ref{QGDLongForces}(b) as a solid line, is the amount
of energy removed each time the fields are switched, $\Delta KE$. We
represent $\Delta KE$ as sum of
sine-functions~\cite{meerakker:053409}
\begin{equation}
\Delta KE(\phi) = \sum_{n=odd}a_n\sin(n\phi), \label{DeltaKEFit}
\end{equation}
where we have used the definition of the phase angle $\phi =
(z/L)180^\circ$. A fit of the first three terms of this equation to
the actual $\Delta KE$ for deceleration stages spaced by 11 mm is
shown as a dashed line in Fig.~\ref{QGDLongForces}, resulting in the
fit values $a_1 = 1.221$ cm$^{-1}$, $a_3 = 0.450$ cm$^{-1}$, and
$a_5 = 0.089$ cm$^{-1}$. Using this fit, we derive the equation of
motion of the molecules about the synchronous molecule position as

\begin{equation}
\frac{d^2\Delta\phi}{dt^2}+\frac{\pi}{mL^2}(\Delta
KE(\Delta\phi+\phi_0)-\Delta KE(\phi_0))=0, \label{QGDN2}
\end{equation}
where we have used the excursion of the molecule from the
synchronous molecule $\Delta\phi = \phi - \phi_o$. The maximum
stable forward excursion of a non-synchronous molecule is exactly
the same as a traditional decelerator and is given
as~\cite{Hudson:2004}
\begin{equation}
\Delta\phi^+_{max}(\phi_o) = 180^\circ - 2\phi_o. \label{QGDPhiMax}
\end{equation}
We calculate the work done in bringing a molecule starting at this
position with zero velocity to the synchronous molecule position as
\begin{widetext}
\begin{equation}
W(\phi_o) = \int_{Start}^{End} F dx =
-\frac{1}{\pi}\int_{\Delta\phi^+_{max}(\phi_o)}^0
\sum_{n=odds}(a_n[\sin(n(\Delta\phi + \phi_o)) -
\sin(n\phi_o)])d(\Delta\phi). \label{QGDWork}
\end{equation}
Integrating this equation and setting it equal to the kinetic energy
yields the maximum stable excursion velocity:
\begin{equation}
\Delta v_{max}(\phi_o) =
2\sqrt{\sum_{n=odds}\frac{a_n}{m\pi}\left(\frac{\cos(n\phi_o)}{n} -
(\frac{\pi}{2} - \phi_o)\sin(\phi_o)\right)},\label{QGDVMax}
\end{equation}
where $\phi_o$ is now in radians.
\end{widetext}

\begin{figure}
\begin{center}
\resizebox{1\columnwidth}{!}{
    \includegraphics{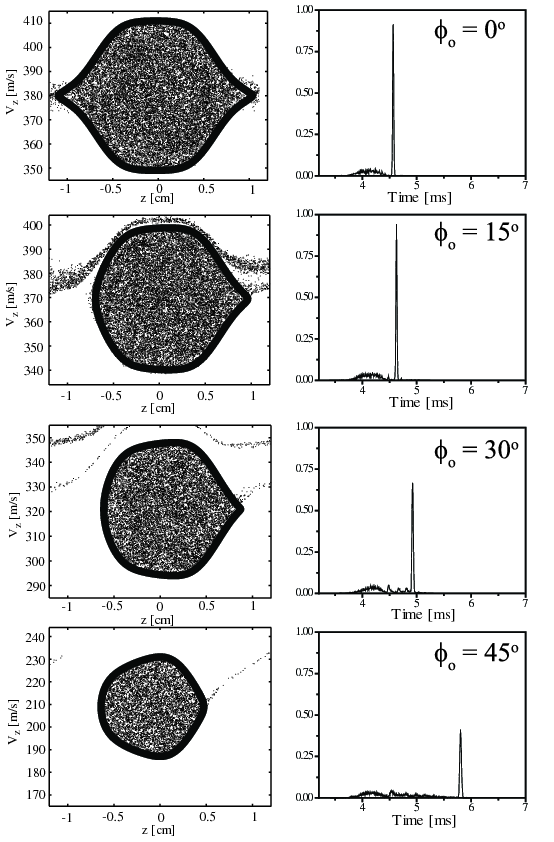}%
}
\end{center}
\caption{The left column is stable phase space of molecules
decelerated inside the QGD. The solid line is the separatrix
predicted by the theory, while the points represent positions of
molecule in the 3D Monte Carlo simulations. The right column shows
the ToF spectra of OH molecules in the $|2\pm2,-\rangle$ state at
the exit of this decelerator which has 142 deceleration stages,
along with 142 quadrupole stages. \label{QGDPSToF}}
\end{figure}

Using Eqs.~\ref{QGDN2}-~\ref{QGDVMax} it is possible to solve for
the longitudinal separatrix, which separates stable deceleration
from unstable motion inside the decelerator. These separatrices are
shown (thick black lines) along with the results of Monte Carlo
simulations of a QGD in the left column of Fig.~\ref{QGDPSToF} for
successive phase angles. The longitudinal phase space is shown with
each dot representing the position of a stable molecule. The lack of
structure inside these separatrices is evidence of the lack of
coupling between the transverse and longitudinal modes.  The right
column, which shows simulated ToF curves, reveals a single stable
peak arriving at later times as $\phi_o$ is increased. These
simulations are for a single $|2,\pm2,-\rangle$ state of OH and do
not exhibit the large background contribution of the other states of
OH present in experimental ToF data.

\begin{figure}
\begin{center}
\resizebox{1\columnwidth}{!}{
    \includegraphics{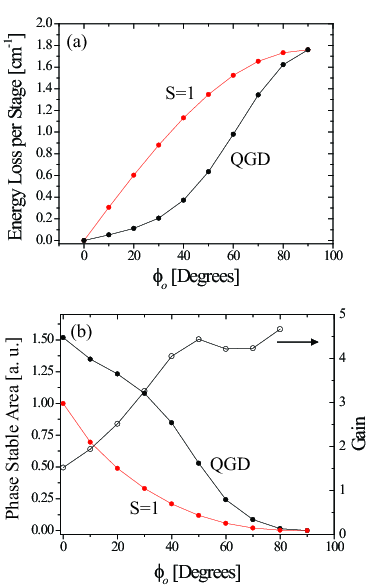}%
}
\end{center}
\caption{(color online) (a) The energy removed per stage as a
function of phase angle for both traditional deceleration and
deceleration with a QGD. Both curves are calculated for OH in the
$|2,\pm2,-\rangle$ state. (b) The calculated phase-stable area for
deceleration versus phase angle for traditional deceleration and
deceleration with a QGD is plotted on the left axis. The gain of the
QGD over traditional deceleration is plotted on the right
axis.\label{QGDPS}}
\end{figure}

From the comparison of the simulations with the analytical results
represented by the separatrices, we see that the simple theory of
Eqs.~\ref{QGDN2}-~\ref{QGDVMax} is quite accurate in describing the
longitudinal performance of the QGD. Thus, by numerically
integrating the area inside these separatrices, we can predict the
longitudinal performance of the QGD relative to the traditional
decelerator. As seen in Fig.~\ref{QGDPS}(a), the energy removed per
stage of the QGD is less steep with $\phi_o$ because the decelerator
stage spacing (in this simulation) is twice that of a traditional
decelerator. For this reason, QGD deceleration with the same
$\phi_o$ as in a traditional decelerator leads to a faster beam,
given the same number of deceleration stages. When comparing the
longitudinal acceptance of the two types of decelerators it is
important to take the limit of high phase angles where both values
of $\Delta KE$ converge. Nonetheless, the QGD shows significant gain
over traditional deceleration as shown in Fig.~\ref{QGDPS}(b). This
gain is primarily due to the increased physical size of the stable
longitudinal phase space resulting from the larger deceleration
stage spacing.

\begin{figure}
\begin{center}
\resizebox{0.8\columnwidth}{!}{
    \includegraphics{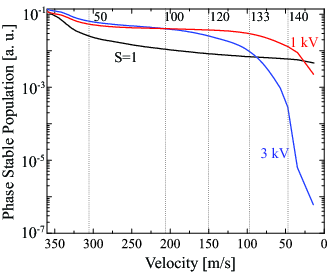}%
}
\end{center}
\caption{(color online) Simulations of the phase-stable molecule
number as a function of stage number in the QGD and $S=1$
decelerator. The voltage on the quadrupole stages of the QGD is held
constant throughout the deceleration sequence. All simulation data
is for $v_{final}=14$ m/s. The traces shown are $S=1$ deceleration
at $\phi_o = 30.43^\circ$, QGD operated with $\pm1$ kV on the
quadrupoles, and QGD operated with $\pm3$ kV on the quadrupoles.
Note the decrease in stable molecule number in the last several
stages for the QGD results. This decrease is due to transverse
overfocusing of the slow molecules through the final few quadrupole
stages, and suggests that scaling of quadrupole voltage is
necessary. \label{QGDSNLoss}}
\end{figure}



\begin{figure}
\begin{center}
\resizebox{0.8\columnwidth}{!}{
    \includegraphics{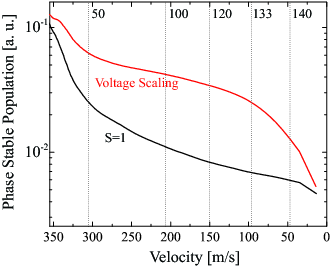}%
}
\end{center}
\caption{(color online) Monte Carlo simulation results for the
decelerated molecule number using traditional $S=1$ deceleration
($\phi_{0}=30.43^{\circ}$) and deceleration using a QGD
($\phi_{0}=52.75^{\circ}$) with a dynamic voltage scaling of
$(v/v_{initial})^{0.875}$. For both curves, 142 stages of
deceleration were used, and $v_{final}=14$ m/s. The different phase
angles chosen for the two decelerators are a result of their
differing potential profiles for deceleration. The vertical dashed
lines represent the deceleration stage at the given velocity. Note
that when the quadrupole voltage within the QGD is scaled in this
manner, we observe a 40\% gain in molecule number at 14 m/s, and a
factor of 5 gain over $S=1$ at higher velocities.
\label{LoadingCurve}}
\end{figure}

It is important to note that since Fig.~\ref{QGDPS} compares only
the total area inside the separatrix--and $S = 1$ deceleration does
not completely fill this area due to the coupling effects--this gain
is actually an underestimate of the QGD longitudinal performance.
Furthermore, these graphs do not include transverse focusing
effects, which can only be properly included through detailed
simulation. The results of Monte Carlo simulations including these
transverse effects are shown in Fig.~\ref{QGDSNLoss}. The number of
decelerated molecules versus final speed is plotted for both the
traditional decelerator operating at $S = 1$ and the QGD operating
at two different quadrupole rod voltages, $\pm1$ kV and $\pm3$ kV.
While the QGD initially delivers more molecules, once the molecules
are decelerated below 100 m/s, the decelerated molecule number falls
off abruptly. This behavior is expected since, for these
simulations, the voltage on the quadrupole-guiding stages is held
constant throughout the slowing sequence. As detailed in Eq. (3) of
Ref.~\cite{Bochinski:2004}, the focal length of a transverse guiding
element is directly proportional to the molecular kinetic energy.
Therefore, as the mean speed of the packet is decreased, the
molecules are overfocused and collide with the decelerator
electrodes. This can be prevented by lowering the voltage on the
quadrupole-guiding stages during the deceleration process.
Figure~\ref{LoadingCurve} displays simulation results of
deceleration with this dynamically scaled voltage compared to $S=1$
slowing at $\phi_o$ = 30.43$^\circ$. For this simulation, the
quadrupole voltages are scaled by $(v/v_{initial})^{0.875}$ after
each deceleration stage, where $v$ is the instantaneous packet
velocity directly following each stage switch. The exponent of 0.875
is found empirically to produce the most gain at $v_{final}=14$ m/s.
For ease of simulation, the transverse forces are scaled by
$(v/v_{initial})^{0.875}$ whenever the molecules are closer to a
quadrupole-guiding stage than to a deceleration stage. While this
may be a poor approximation at the lowest speeds, it will likely
lead to an underestimate of the decelerated molecule number since
the transverse guidance of the quadrupole-guiding stage extends
beyond this regime. Even if it leads to an overestimate, proper
control of the quadrupole voltages may compensate any overfocusing
introduced by the decelerating elements. As seen in
Fig.~\ref{LoadingCurve}, dynamically controlling the voltage of the
quadrupole-guiding stages leads to a factor of 5 increase in
decelerated number for larger velocities ($>$80 m/s) and delivers
about 40\% more decelerated molecules than traditional $S=1$
deceleration provides at the lowest final speeds (14 m/s). Because
the voltages applied to the quadrupole-guiding stages are relatively
low, dynamic control of them should be possible using an analog
waveform generator and high-voltage amplifier. It should be noted
that the optimal voltage scaling may vary among decelerators since
the real focal length depends sensitively on the electrode
construction, and at low speeds the transverse focusing of the
decelerator electrodes becomes significant. This is, presumably,
because the overfocusing of the decelerator electrodes can be
compensated to a certain degree by injecting molecules which are
already slightly overfocused. In other words, two focusing stages
can overcome the overfocusing of a single stage. Thus, it may be
possible to use adaptive algorithms to optimize the quadrupole
voltage or change the design of the decelerating electrodes so that
they provide less transverse focusing, and maximize the number of
decelerated molecules beyond what is reported here~\cite{Joop:2006}.

\begin{figure}
\begin{center}
\resizebox{0.8\columnwidth}{!}{
    \includegraphics{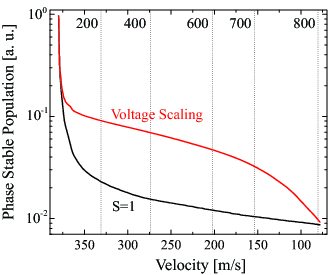}%
}
\end{center}
\caption{(color online) $S=1$ simulation results for $v_{final}=14$
m/s ($\phi_{0}=5.22^{\circ}$) plotted along with simulation results
using the voltage-scaled QGD to the same $v_{final}$
($\phi_{0}=23.5^{\circ}$). The number labeling each vertical dashed
line is the number of deceleration stages necessary to reach the
given velocity. Note the large number of stages (803) used to reach
14 m/s, which suggests that a very long QGD may be employed for
slowing molecules with a poor Stark shift to mass ratio.
\label{QGDSNLoading}}
\end{figure}

One important advantage of the QGD over traditional decelerators is
its inherent ability to support more deceleration stages. Because
the molecules experience a tunable transverse focusing element after
each deceleration stage, there is little loss in efficiency by
extending the number of deceleration stages. In fact, as seen in
Fig.~\ref{QGDSNLoading}, there is essentially no loss until the
molecules are decelerated to the lowest speeds previously discussed.
This low-velocity loss is due to the aforementioned transverse
overfocusing and longitudinal reflection. The former loss mechanism
may be overcome, while the latter presents a fundamental limit. Even
with this loss, the QGD outperforms both $S = 3$
(Fig.~\ref{S=3StageNumber}) and $S = 3+$ (Fig.~\ref{S=3+Number}).
Thus, the QGD is an ideal decelerator for more efficiently producing
cold molecules and, perhaps more importantly, the ideal decelerator
for slowing molecules with poor Stark shift to mass ratios, like
H$_2$O and SO$_2$.

\section{Conclusion}

In summary, we identify several loss mechanisms observed in the
operation of Stark decelerators and perform initial experiments and
detailed Monte Carlo simulations to address them. While the use of
decelerator overtones yields improvement over $S=1$ deceleration at
high to intermediate speeds ($v_{final}>80$ m/s), the loss at very
low velocities remains problematic. The QGD solves the problem of
coupling between the transverse and longitudinal motions inside a
Stark decelerator by introducing independent transverse focusing
elements. By dynamically controlling the focal length (voltage) of
these guiding elements, large improvements (factor of 5) in
deceleration efficiency can be achieved for $v_{final}>80$ m/s, and
gain of $\sim40\%$ is predicted for the lowest velocities.
Furthermore, it appears that with dynamic control of the guiding
stage focal length there should be no limit to the length of
decelerator that can be built. This enables the deceleration of
molecules with a poor Stark shift to mass ratio. However, we note
that none of the techniques described in this Article mitigate the
longitudinal low-velocity loss due to reflection, which appears to
be a fundamental component of Stark deceleration. Building upon the
strong correlation between simulation and experimental results, we
are confident that the simulations presented in this work provide a
solid guideline for future implementations of Stark deceleration.

\begin{acknowledgments}
We wish to thank H. L. Bethlem and S.Y.T. van de Meerakker for
useful discussions. We also thank M. Yeo, H. Lewandowski, and D.
Nesbitt for reading the manuscript. This work is supported by DOE,
NIST, and NSF. B. L. Lev is an NRC postdoctoral fellow.
\end{acknowledgments}


\end{document}